# X-ray Diffraction and Molecular Dynamics Study of Medium-range Order in Ambient and Hot Water

Congcong Huang[1], K. T. Wikfeldt[2], D. Nordlund[1], U. Bergmann[3], T. McQueen[1], J. Sellberg[1,2], L. G. M. Pettersson[2] and A. Nilsson[1,2]

[1]Stanford Synchrotron Radiation Lightsource, P.O.B 20450, Stanford, CA 94309, USA

[2]Department of Physics, AlbaNova, Stockholm University, S-106 91 Stockholm, Sweden

[3] Linac Coherent Light Source, SLAC National Accelerator Laboratory, P.O. Box 20450, Stanford, CA 94309, USA

**Abstract**

We have developed x-ray diffraction measurements with high energy-resolution and accuracy to study water structure at three different temperatures (7, 25 and 66 °C) under normal pressure. Using a spherically curved Ge crystal an energy resolution better than 15 eV has been achieved which eliminates influence from Compton scattering. The high quality of the data allows a precise oxygen-oxygen pair correlation function (PCF) to be directly derived from the Fourier transform of the experimental data resolving shell structure out to ~12 Å, *i.e.* 5 hydration shells. Large-scale molecular dynamics (MD) simulations using the TIP4P/2005 force-field reproduce excellently the experimental shell-structure in the range 4-12 Å although less agreement is seen for the first peak in the PCF. The Local Structure Index [J. Chem. Phys. **104**, 7671 (1996)] identifies a tetrahedral minority giving the intermediate-range oscillations in the PCF and a disordered majority providing a more featureless background in this range. The current study supports the proposal that the structure of liquid water, even at high temperatures, can be described in terms of a two-state fluctuation model involving local structures related to the high-density and low-density forms of liquid water postulated in the liquid-liquid phase transition hypothesis.

**Introduction**

Despite the importance of water to our daily life, the structure of the hydrogen-bond (H-bond) network of liquid water at ambient conditions has been debated since the 1930s. Surprisingly, this has so far not resulted in a coherent physical picture, in part because of the difficulty in obtaining detailed direct experimental evidence on the three-dimensional (3D) arrangement of H-bonds with its dynamics (breaking and reforming) on a time scale of picoseconds. In contrast to water under ambient conditions, the structure of supercooled water (a metastable form of water below 0 °C) is commonly accepted to exhibit fluctuations between at least two distinct structural states with different densities, referred to in the literature as low-density liquid (LDL) and high-density liquid (HDL), respectively.

Experimentally, only indirect evidence on the coexistence of LDL and HDL in liquid water has been found where, *e.g.*, Angell *et al.* [1, 2] first observed a power-law divergence of thermodynamic properties of water upon approaching a singular but experimentally inaccessible temperature of 228 K at ambient pressure. Mishima and Stanley [3] observed indications of a first-order liquid-liquid phase transition during melting of high-pressure phases of ice. Liu *et al.* [4] observed a density minimum of deeply supercooled water in a nano-confined geometry and very recently a density hysteresis was reported for supercooled nanoconfined heavy water [5]. Bosio *et al.* [6, 7] and Huang *et al.* [8, 9] found enhanced density fluctuations associated with anomalous scattering intensity at low momentum transfer upon water cooling. Bellissent-Funel[10] interpreted neutron diffraction data on water in terms of two structural limits connected to low-density amorphous (LDA) and high-density amorphous (HDA) ice, respectively, and proposed a "two-level"-type model of water. Soper and Ricci [11] uncovered the structure correlations of possible HDL and LDL via a series of high-pressure neutron diffraction measurements.

Such a two-state model in the supercooled regime finds support from many molecular dynamics (MD) simulations, which can investigate lower temperatures than possible experimentally by avoiding the homogeneous nucleation occurring in real bulk water. While the main contesting thermodynamic scenarios are the liquid-liquid critical-point (LLCP) [12], the singularity-free (SF) [13], the critical-point free (CPF) [14] and the stability-limit (SL) hypothesis [15], most molecular dynamics (MD) force-fields seem to exhibit one or more critical points in the deeply supercooled liquid region, *e.g.*, refs. [12, 16-20], and thus support the LLCP scenario. All four scenarios were unified in a single Hamiltonian cell-model [21] where two key parameters of molecular interactions in water, the cooperative and the directional components of the H-bond interaction, could be tuned to obtain phase diagrams corresponding to each scenario. For the largest range and most realistic parameter values, however, the LLCP scenario with a second critical point at positive pressures was obtained.

Experimental signatures of anomalous behavior related to the proposed coexistence of LDL and HDL generally become indistinct for bulk water under ambient conditions compared to those in the supercooled regime. However, recent x-ray spectroscopy studies on ambient water have been interpreted in terms of two distinct H-bond structural motifs with a dominant HDL-like species [9, 22-26] although this interpretation is still debated (see discussion in refs. [22, 23, 27]). High accuracy small-angle x-ray scattering (SAXS) measurements of water in the ambient regime were interpreted in terms of density fluctuations caused by a combination of structural fluctuations (increasing with *decreasing* temperature) and normal stochastic number fluctuations (increasing with *increasing* temperature) where the latter dominate at elevated temperatures [9]. This interpretation was contested in a comment by Soper *et al.* [28, 29] and Clark *et al.* [30], but subsequent SAXS measurements [8] and simulations [31] extending the temperature range to encompass both ambient and supercooled water have corroborated the original interpretation. The onset of a fractional Stokes-Einstein relation below 290 K has furthermore been connected to a structural transition from predominantly tetrahedral, LDL-like at supercooled conditions to predominantly HDL-like with significantly weakened H-bond network at ambient conditions [32].

The water radial distribution functions (RDF) derived directly from diffraction measurements play an important role in this long-standing debate, not only since the peak positions and amplitudes give information on water structure, but also for benchmarking force-fields for simulations of water. As one of the most direct measurements, x-ray diffraction (XRD) supplies insight into the nature of water structure via the two-body oxygen-oxygen pair correlation function (O-O PCF). This has been frequently measured dating back to the 1930s, *e.g.*, by Morgan and Warren [33], Narten and Levy [34, 35], Okhulkov *et al.*[36], Hura *et al.* [37-39], Tomberli *et al.* [40], Yokoyama *et al.* [41], Fu *et al.* [42] and Neuefeind *et al.* [43] to just name a few. However, there is no uniform conclusion yet, largely due to conflicting results for the height and profile of the first PCF peak and also regarding fine structures on the high-*r* side of the first PCF peak at about 3.5 Å (see discussion in ref. [42]). Furthermore, structure beyond the 3$^{rd}$ solvation shell has rarely been resolved for ambient water. The reason is that an accurate XRD measurement is hindered by uncertainties from experiment and data analysis, such as the large scattering background from the sample container when such is used (see Fig. 3 in ref. [39]), the low signal-to-noise ratio of water scattering, the separation of elastic and Compton scattering and the choice of coherent self-scattering factor based on atomic or molecular scattering.

Temperature-dependent changes in the O-O PCF provide information on structural changes in the liquid as function of temperature. In the early work by Narten and coworkers[44] a decrease in the $2^{nd}$ shell correlation at 4.5 Å and increasing asymmetry in the first peak were observed upon heating. Urquidi *et al*. [45, 46] pointed out the similarity between pressure- and temperature-induced changes in the O-O PCF based on data from Okhulkov *et al*.[36] and Bosio *et al*. [47]. They specifically noted that with increased temperature or pressure the correlation in the first minimum at 3.5 Å increased with concomitant decrease around 4.5 Å, and a shift in peak position with pressure in the region of 6.5-7 Å was also noted. Here, we revisit this old but still unsolved question, the structure of ambient water, by performing temperature-dependent XRD with a high energy resolution and accuracy. In the present experiment we use the traditional angular-dispersed setup with a monochromatic beam from a bright synchrotron radiation source. A combination of enhanced signal-to-noise ratio and higher energy resolution, better than 15 eV at 17,000 eV photon energy, is achieved through a container-free water-jet sample and a spherically curved single-crystal to, directly in the experimental setup, separate out the Compton scattering from the pure elastic scattering . We verify and extend previous diffraction studies and use large-scale MD simulations to investigate the structural origin of features detected in the experimental PCFs up to ~12 Å. We find that the two-state fluctuation model, which has been widely applied to describe water in the supercooled regime, may be appropriate also at ambient conditions since the presence of fine features in the PCFs out to around 12 Å together with the observation of a broad and asymmetrical first O-O peak is hardly compatible with a continuum model description. Indeed, our MD simulations confirm that the intermediate-distance correlations derive from tetrahedral LDL-like species, while a low and asymmetrical first peak is a signature of disordered HDL-like species.

**Methods**

*Experiment*

The experiments were performed at beamline 7-2 at Stanford Synchrotron Radiation Lightsource (SSRL). A Huber 6-circle diffractometer was used with a monochromatic beam of 17,000 eV. A water jet with a diameter of 360 μm was aligned at the rotation center of the diffractometer and kept in a helium environment to reduce the air scattering signal [42]. The water sample was circulated through a pumping system with temperature control. No container scattering contributes to the signal and potential beam damage is also avoided through the flow system. The diffraction scans were carried out in an angular-dispersive setup in the momentum transfer $q$ range of 0.5 to 16.0 Å$^{-1}$, where $q$ is defined as $q=4\pi sin\theta/\lambda$ ($\lambda=0.73$ Å is the incident wave length and $2\theta$ is the scattering angle as indicated in Fig.1). The data were taken in multiple diffraction scans with an even $q$ step size of 0.1 Å$^{-1}$. A constant dose model was applied in order to increase the statistics at large $q$ where the elastic scattering signal is about 30 times weaker than that at $q=2$ Å$^{-1}$. In total, at each $q$ point there are about 85,000 photon counts for each temperature measurement (7˚, 25˚ and 66 ˚C) and the statistical uncertainty is thus below 1%.

Asides from the weak scattering of water, the energy tail of Compton scattering at intermediate $q$ range overlapping with the elastic signal is the potentially greatest source of systematic error that must be considered in the experimental design. A Germanium single-crystal analyzer was chosen to achieve high energy resolution. As illustrated in Fig.1, the crystal is mounted in a Rowland geometry to refocus the scattered beam on a photo multiplier tube (PMT) detector, which is shielded by a beamstop. We use Ge(880) reflection at a Bragg angle of 46.8˚ and focus distance of 36.4 cm. The energy resolution was measured by rocking the analyzer crystal at selected $2\theta$ points and the result is plotted in Fig. 2a as a function of $q$. The intrinsic energy band width of the Ge crystal arising from its finite size (10 cm in diameter) is almost negligible (<1 eV). The main limitation of the energy resolution is given by the inherent energy band width of the incoming beam (~10 eV) as well as by the finite size of the water jet. As a consequence, the resulting energy resolution is better than 15 eV at scattering angles away from $2\theta=90˚$ as shown in Fig. 2a, which represents a significant improvement compared to previous experiments. For example, Hura et al. [37] used a charge-coupled device area detector with a purely theoretical correction for Compton scattering, Badyal et al. [48] used an energy dispersive detector with 400 eV resolution and Fu et al. [42] used a graphite diffracted analyzer with a resolution of 50 eV and therefore fitting procedures had to be used in their data analysis to extract the elastic scattering signal. In addition, we checked the signal level of inelastic scattering at ~30 eV away from the elastic peak by offsetting the crystal angle by 0.1° from its Bragg condition. As shown in Fig. 2b, it is clear that inelastic scattering contributes less than 1% of the total signal in the low $q$ range (<4 Å$^{-1}$), whilst it completely disappears at large $q$ as a consequence of zero overlap in energy between elastic and inelastic scattering when $q$ increases. We would thus expect an even smaller influence with 15 eV energy resolution and corrections for Compton scattering are therefore not needed for the later data analysis.

A $q$ resolution of about 0.1 Å$^{-1}$ was realized by placing an entrance slit between water jet and analyzer crystal with an in-plane opening of 7 mm as illustrated in Fig 1. It was found that an out-of-plane opening of the entrance slit only has a secondary effect on the resolution, $\Delta q$, thus we left it open to see the whole crystal. An exit slit directly prior to the detector was also used in order to further reduce scattering background given the fact that the in-plane scattered beam is focused on the detector according to the Rowland geometry. Three experimental corrections were performed to extract the elastic signal from the raw scattering intensities. They are scattering background correction (mainly from air scattering and detector dark signal), beam polarization correction (from a linearly polarized incident beam with ~5% fraction of a vertical component) and optical aberration of the analyzer crystal (due to the finite size of the water jet and in-plane beam focus). The re-scale factors corresponding to these three corrections and the modified raw data on a logarithmic scale are shown in Fig 3. It is clear that the interference oscillations of water scattering are visible up to $q$=16.0 Å$^{-1}$ in the raw data and the corrections supply a structure-less envelope function which only contributes to the intramolecular signal ($r$<2.5 Å) as discussed below. We also note that the correction accounting for multiple scattering effects is negligible in our current experiment due to the small sample dimension.

*Molecular Dynamics (MD) Simulations*

We perform classical MD simulations using the TIP4P/2005 force-field [49] with 45,000 molecules in the constant pressure, constant temperature (NPT) ensemble. The pressure is constrained to 1 bar using the Parrinello-Rahman[50] barostat and different temperatures are obtained using the Nosé-Hoover[51, 52] thermostat. The equations of motion are integrated using the leap-frog algorithm with a 2 fs timestep. Long-range electrostatic interactions are treated using the particle-mesh Ewald method and long-range dispersion corrections are applied for the truncated Lennard-Jones interaction. Intramolecular geometries are constrained using the LINCS algorithm, and the simulations were run on a parallel platform using the Gromacs package [53].

We analyze structural order and disorder using a parameter called the local-structure-index (LSI) [54, 55] defined for each molecule $i$ by ordering the nearest neighbors $j$ according to increasing distance to molecule $i$ as $r_1 < r_2 < r_3 < \cdots < r_{n(i)} < 3.7$ Å $< r_{n(i)+1}$ where $n(i)$ is the number of molecules that are within 3.7 Å from molecule $i$. The LSI distinguishes molecules with well separated first and second coordination shells from molecules with disordered environment that contains neighboring molecules in interstitial positions through the index $I(i)$ defined as

$$I(i) = \frac{1}{n(i)} \sum_{j=1}^{n(i)} [\Delta(j;i) - \overline{\Delta}(i)]^2, \qquad (1)$$

where $\Delta(j;i) = r_{j+1} - r_j$ and $\overline{\Delta}(i)$ is the average of $\Delta(j;i)$ over all neighbors $j$ of molecule $i$. A low-LSI corresponds to a disordered local environment while a high-LSI indicates a highly structured, tetrahedral coordination.

**Results**

**I Comparison of different temperatures**

The water structure is analyzed based upon a per-molecule basis by using a quantum mechanically calculated molecular scattering factor (MSF), $F(q)^2$, of an isolated water molecule [56]. The elastic scattering intensity $I(q)$ was first normalized with respect to $F(q)^2$ since we have data extending to a sufficiently large maximum $q$ where it is expected to oscillate around the MSF with a damped amplitude. The molecular structure factor $S(q)$ was then derived from the normalized scattering intensity $I(q)$ as $I(q) = F(q)^2 S(q) + F(q)^2$ under the spherical-molecule approximation. The resulting $S(q)$ at three temperatures are shown in Fig 4. It can be seen that the periodicity of interference oscillations decreases slightly as temperature increases from 7°C to 66 °C. Moreover, the doublet of the first $S(q)$ maximum near 2-3 Å$^{-1}$ becomes better resolved as temperature decreases. Note that oscillations in S($q$) are observed all the way out to 16 Å$^{-1}$.

The PCF, g(r), and RDF, $4\pi r^2\rho_0 g(r)$, are widely applied concepts in structural analysis. The former describes the probability of finding a particle at distance $r$ from another particle, where the orientation of particles is averaged over angles. The current work deals only with the distribution of distances related to the center of electron density of water molecules, which to a good approximation coincides with the oxygen atoms. By Fourier transforming the experimental data, the RDF can be obtained from the structure factor $S(q)$ through the following relationship

$$4\pi r^2 \rho(r) = 4\pi r^2 \rho_0 g(r) = 4\pi r^2 \rho_0 + \frac{2r}{\pi}\int_0^{q_{max}} e^{-\alpha q^2} q S(q)\sin(qr)dq, \qquad (2)$$

where $\rho_0$ is the average molecular density of water in the present case and $\rho(r)$ is the average local density at a distance $r$ from the average center. An exponentially decaying window function, $e^{-\alpha q^2}$ with α=0.004, was used in order to decrease the magnitude of spurious ripples resulting from the truncation errors associated with the Fourier transform [57]. We note that the spurious ripples are strongly reduced but not completely eliminated by this method. Generally, the more damped the $S(q)$ interference oscillations are at the cut-off $q_{max}$, the less influence the Fourier truncation will have on the PCF. We also note that possible normalization errors are found to contribute exclusively to the intramolecular distances, *i.e.* $r$<2.4 Å, leaving the PCF at intermolecular distances almost unaffected as demonstrated in ref. [42].

The derived O-O PCFs comparing the three temperatures from the current experiment are shown in Fig. 5. The first PCF peak, associated with the short-range order (SRO) around the nearest-neighbor distance in water, is observed to shift outwards with temperature from 2.81 Å at 7 ˚C, 2.82 Å at 25 ˚C and to 2.84 Å at 66 ˚C. Such a peak shift is mainly attributable to the normal thermal expansion, corresponding to the shortening of the interference oscillation periodicity of $S(q)$ with increasing temperature. The second shell is centered at 4.5 Å, satisfying the relationship of $\sqrt{8/3}$ times the first PCF peak which indicates the existence of configurations with a local water structure close to tetrahedral. On close inspection, we observe that the first PCF peak profile becomes more asymmetric with increasing temperature, leading to intensity "leakage" to the so-called interstitial distances, *i.e.* $r$~3.5 Å. Although the possibility of the existence of a distinct interstitial peak in the water PCF at elevated temperatures cannot be concluded from the current measurement due to the truncation errors, the asymmetric broadening of the first PCF peak towards the larger *r* side is rather well determined. A similar increased intensity at interstitial distances has previously been observed from isochoric temperature pairs in $D_2O$ [47] and $H_2O$ under high pressures [36] and discussed in terms of an outer structure two-state mixture model by Robinson and coworkers[45, 46]. The normal thermal expansion cannot by itself explain this observation, implying that a local structure different from ice-like tetrahedral evolves with increasing temperature and exhibits a longer O-O nearest-neighbor distance.

Our temperature-dependent O-O PCF at small *r* is in good agreement with the early XRD data of Narten *et al*, [44] which covered a wider temperature range, where an asymmetric broadening of the first peak and a reduced second peak height were observed upon heating. Our room temperature PCF is also consistent with that of Fu *et al.* [42] as well as of Neuefeind *et al.*[43] which both allow a similar Fourier transform approach applied to raw scattering data without a pre-defined model. On close inspection our data do not support the fine-structure reported at 3.4 Å by Fu *et al.* [42] which we speculate was rather due to Fourier truncation effects due to the limited *q*-range of observed oscillations in their data ($q<13$ Å$^{-1}$); however, consistent with their conclusions we find enhanced intensity in the interstitial region, albeit no sharp feature. On the other hand, the current result gives a much lower and wider first O-O PCF peak compared to that of Hura *et al.* [37-39] which was obtained by fitting the total scattering *I(q)* in terms of *I(q)* from a basis set of PCFs obtained from various MD simulations and experimental data. The agreement at the time with the independent analysis by Soper [58] of neutron diffraction data using the empirical potential structure refinement (EPSR) procedure[59, 60] with SPC/E as initial force-field was taken as indication that a correct solution had been arrived at. We speculate, however, that the inconsistency of these studies compared with the early work by Narten and Levy[35, 61] as well as with the present data and recent analyses [42, 43, 62-64] may stem from the involvement of MD force-fields, whose O-O peak heights, widths and positions have only in recent years become questioned due to poor agreement with experimental scattering data [42, 43, 62-64].

The PCF at intermediate distances is magnified by plotting in Fig. 6a the scaled difference in the radial distribution function (dRDF) defined as $4\pi r^2 \rho_0 (g(r)-1)$. We observe structural correlations up to *r*~12 Å, indicating the presence of a medium-range order (MRO) in the liquid. In particular, the 4$^{th}$ PCF peak at *r*~9 Å and the 5$^{th}$ peak at *r*~11 Å are resolved here for the first time from XRD measurements for ambient and hot water. After the 5$^{th}$ shell, the correlations are gradually washed out within the noise level of the experiment. A 5$^{th}$ PCF peak, similar to the present data, has been observed in supercooled water in two previous independent x-ray studies [41, 43]. Yokoyama *et al.*[41] studied both supercooled and ambient water, but the signal-to-noise level made it difficult to draw firm conclusions on intermediate-range correlations at 25 °C while the more prominent shell structure at supercooled temperatures was proposed to imply a clathrate-like structure mainly made of water pentagons. It is also interesting to observe the different temperature dependence in the different O-O PCF peaks as shown in Fig. 6a: the 1$^{st}$ and 4$^{th}$ peaks exhibit less temperature sensitivity compared to the 2$^{nd}$, 3$^{rd}$ and 5$^{th}$ peaks whose magnitudes strongly increase as temperature decreases from 66 °C to 7 °C. It directly indicates that there are temperature-dependent structural changes in liquid water in addition to the effects of disorder induced by normal thermal motion.

**II Comparison to molecular dynamics simulations**

The TIP4P/2005 force-field[49] has been demonstrated to reproduce the minimum in isothermal compressibility, maximum in density and furthermore gives a good description of the crystalline phases[65, 66]. This model also shows an enhancement in the structure factor at low $q$ giving a near-quantitative agreement with small angle x-ray scattering data [31, 67]. It becomes clear by inspecting Fig. 6b, however, that the force-field overestimates the local structure of liquid water, revealed by a much narrower and higher first peak in the PCF at a shorter nearest neighbor distance compared to the experimental data shown in Fig 6a. Moreover, the asymmetric broadening of the first PCF peak is not reproduced well by the simulation and the interstitial distances lack intensity compared to the experiment. Although the first-shell structure is too ordered in this model, it gives an overall good agreement with experimental small-angle scattering data and thermodynamic properties [31, 49, 65]. We therefore use this model to provide further insights into the intermediate-range correlations observed in the current experiment.

As shown in Fig. 6b, the TIP4P/2005 simulation clearly contains the intermediate-range correlations giving the 4$^{th}$ and 5$^{th}$ PCF peaks in the correct positions. It is difficult however to directly compare the peak magnitudes and widths because of the enhanced noise level at large $r$ in the PCF derived from experiment. In terms of the temperature dependence, an excellent agreement between the current MD simulation and experimental data is obtained: the 5$^{th}$ peak at $r$~11 Å is observed to significantly increase in amplitude with decreasing temperature both in the XRD data and in the TIP4P/2005 simulation, while the amplitude of the 4$^{th}$ peak exhibits less dependence on temperature. Moreover, the position of the 4$^{th}$ peak is seen to shift to larger distances at higher temperatures as indicated by the dashed line in Fig. 6b, consistent with the shift observed between 7 and 66 °C in the experimental data shown in Fig. 6a. To investigate the structural origin of these peaks at intermediate distances we characterize the molecules in the simulation according to the LSI parameter, $I(i)$, and thus define sub-ensembles of water molecules in either disordered or structured environments.

A cut-off value for $I(i)$ to classify all molecules into two classes was taken to be $I_c$=0.03 Å$^2$; molecules with $I(i)>I_c$ are highly structured (LDL-like), while those with $I(i)<I_c$ correspond to relatively disordered structures (HDL-like). Resulting relative populations of LDL-like species using $I_c$=0.03 Å$^2$ were 49%, 44% and 38% at 278 K, 298 K and 340 K, respectively. We find that different choices of the cut-off value do not change the trends discussed here, *e.g.*, using a higher value results in a smaller, but structurally more well-defined, fraction of LDL-like species. For the present purpose we have selected a cutoff that gives similar fractions above and below threshold. The left three panels of Fig. 7 show the decomposed O-O PCFs for the first three hydration shells while the panels to the right focus on the decomposed O-O dRDFs in the region of the 4$^{th}$ and 5$^{th}$ hydration shells, where each component reflects the environment around the respective species (*i.e.* including intra-species and inter-species correlations), as obtained for the two sub-species of TIP4P/2005 water at different temperatures. Note that the decomposed dRDFs have been scaled by the relative fractions of low-LSI or high-LSI species. As seen by the decomposed PCFs, the high-LSI species are characterized by very well defined first and second coordination shells and a deep first minimum, in sharp contrast to the low-LSI species which feature a collapsed second coordination shell and a pronounced shoulder at interstitial distances around 3.5 Å, similar to what is observed for high density amorphous (HDA) ice[68]. It indicates that the shoulder feature at interstitial distances observed in the experimental PCFs (Fig. 5) is exclusively attributable to HDL-like species as characterized by low LSI values in the simulation. Moreover, the 3$^{rd}$ shell around 6.7 Å is shifted to lower distances for low-LSI species, similarly to pressurized water [11, 36, 69]. To a rather large extent, the PCFs of high-LSI and low-LSI species in the present TIP4P/2005 MD simulation thus resemble the PCFs of LDL and HDL respectively, which were derived experimentally from a series of high pressure neutron diffraction measurements [11].

For the intermediate-distance features in the dRDFs shown in Fig.7 we observe that high-LSI species (LDL-like) display two well-defined peaks around 8.7 and 11.0 Å, close to where the 4$^{th}$ and 5$^{th}$ peaks are observed experimentally as shown in Fig. 6a. Both peaks decrease rapidly in amplitude at higher temperatures in part due to the decreasing population of high-LSI species. On the other hand, low-LSI species (HDL-like) exhibit a very different behavior where a rather broad plateau centered around 9.5 Å at 298 K gradually develops into a peak at the *highest* temperature, 340 K. This unusual temperature dependence is responsible for the shift of the 4$^{th}$ PCF peak to larger distances at higher temperatures and the weaker temperature dependence of its amplitude compared to that of the 5$^{th}$ peak as observed in Fig. 6b; the gain in intensity of the low-LSI species with increasing temperature in the region of the 4$^{th}$ peak compensates the loss of contributions from high-LSI species explaining the apparent weak temperature dependence of this peak while, in contrast, both the 3$^{rd}$ and 5$^{th}$ peaks loose amplitude with increasing temperature.

The close similarity between simulated and experimental total PCFs on the intermediate length scale suggests that the 4$^{th}$ and 5$^{th}$ correlation peaks, and in particular the latter, observed in the current XRD study are attributable to the existence of highly ordered (LDL-like) structural environments also in real ambient water, as was also concluded in a previous SAXS study [9]. The loss of intensity in the 5$^{th}$ peak with increasing temperature, as shown experimentally in Fig. 6a and closely reproduced by the simulations in Fig. 6b, can then be regarded as a sign of a conversion of LDL-like to HDL-like species upon heating alongside the increased thermal disorder, as reflected also in simulations by the diminishing contribution from high-LSI species at higher temperatures seen in Fig.7. At a high temperature of 66 ˚C, the persistence of the 4$^{th}$ peak and its shift to larger *r* reflects on the other hand a structural change of disordered HDL-like species where a peak close to 9.3 Å develops with temperature.

Comparing to the present experimental PCF, it seems that both the largely asymmetric 1$^{st}$ peak at 2.8 Å and distinct 4$^{th}$ and 5$^{th}$ peaks at intermediate distances are the key features to interpret the structure of ambient water by utilizing the TIP4P/2005 MD simulations. It is clear that the HDL-like species in the simulation give a distinct interstitial shoulder around 3.5 Å and an asymmetric shape of the 1$^{st}$ PCF peak, the latter in qualitative agreement with the experimental total PCF, thus indicating that HDL-like species contribute to the skewness of the first peak. The LDL-like species, on the contrary, give rise to no intensity at interstitial distances but are characterized by well defined 4$^{th}$ and 5$^{th}$ peaks at around 8.7 and 11 Å. In order to match both the short- and medium-range order observed for ambient and hot water as shown in Fig. 6a, therefore, contributions from both HDL- and LDL-like species, as suggested by the analysis of the MD simulations, seem essential also to describe real water. Furthermore, the opposite temperature-dependence between these two $r$ regions, *i.e.* increased intensity at the interstitial distance upon heating and enhanced 5$^{th}$ PCF peak upon cooling, is in accordance with general expectation for the ratio between HDL- and LDL-like species in supercooled water but here this concept seems to apply also to the ambient regime as suggested both from experiment and the analysis of the present MD simulations.

**Discussion and Conclusions**

A new XRD setup with high energy resolution and container-free sample environment was used to study the structure of liquid water at ambient conditions. The O-O PCF was derived directly from the Fourier transform of the scattering structure factor, separated from the Compton scattering experimentally and with the coherent self-scattering contribution eliminated based on theoretical molecular form factors. The resulting O-O PCF shows an asymmetric first correlation peak with its peak position shifting from 2.81 Å at 7 ˚C, 2.82 Å at 25 ˚C, to 2.84 Å at 66 ˚C (Fig. 5). The peak profile is observed to become more asymmetric with extra intensity appearing at the interstitial distance of $r$~3.5 Å as temperature increases. At intermediate intermolecular distances, the high quality of the current XRD data reveals the existence of medium-range order in ambient and hot water resolving up to the fifth coordination shell (Fig. 6a), which has only been resolved earlier in supercooled water [41, 43]. In comparison, MD simulations using the TIP4P/2005 water model give a much sharper first O-O PCF peak and its outwards shift and asymmetric broadening with increasing temperature is not well reproduced, but the fourth and fifth O-O PCF peaks at 9 and 11 Å are in good agreement with the experimental data including their distinctly different behavior upon changes in temperature. Decomposing the simulated PCF into contributions from different structural species revealed that a sub-ensemble of molecules with a very well defined coordination shell (LDL-like species), as quantified by the LSI parameter, gives these resolved peaks at intermediate distances. These observations from MD simulations validate a conceptual approach in which the experimental XRD results are connected to structurally defined subspecies appearing in the simulations.

In the case of ambient and hot water studied here, our XRD data in combination with structural analyses of MD simulations support the coexistence of two different local structures (HDL-like and LDL-like) also in ambient and hot water. This conclusion is mainly based on the following observations: (i) the non-uniform temperature dependence of the different PCF peaks from both experimental and simulated data (Fig. 6), and (ii) the good agreement between experiment and simulation in the PCFs in the range 4-12 Å, which gives confidence to extract additional information on the PCFs from the simulation, where the LSI parameter applied to the TIP4P/2005 simulated PCF allows a differentiation and description of most PCF features in terms of two different structural contributions (Fig. 7). On the contrary, it is hard to envision a continuum model description of water structure which simultaneously gives a broad and asymmetric peak around the nearest-neighbor distance and well-defined correlation peaks at distances even beyond 1 nm associated with the presence of highly ordered species. Indeed, the large structural differences between sub-species revealed in the decomposed PCFs from the TIP4P/2005 simulations are by themselves a strong indicator that a continuum model description is insufficient for liquid water.

It is also worth noting that the possible existence of structural fluctuations between LDL- and HDL-like environments in water cannot be regarded as concentration fluctuations[28] but rather as the *origin* of the enhanced number density fluctuations appearing upon cooling liquid water as observed through enhanced zero-angle scattering intensity [6-9, 29]; application of pressure furthermore reduces the small-angle enhancement as expected from a picture of the anomalously increased compressibility arising from HDL to LDL fluctuations [70]. As demonstrated in a recent TIP4P/2005 MD simulation study of small-angle x-ray scattering (SAXS) [31], the isothermal compressibility of water derived from $S(0)$ is in excellent agreement with that derived from the fluctuation formula in the NPT ensemble even at deeply supercooled temperatures where HDL-LDL fluctuations are clearly present in the simulations.

Finally, despite the success of applying the TIP4P/2005 MD simulation to reproduce the medium range order observed in ambient and hot water, we notice that there is discrepancy between simulated and experimental PCF, especially around the nearest-neighbor and interstitial distances (Fig. 6). Indeed, such a discrepancy has been widely observed in various classical force-fields and *ab initio* MD simulations. We will address this issue in a forthcoming paper.

**Acknowledgements**

We acknowledge the National Science Foundation (U.S.) (Grant No. CHE-0809324), Basic Energy Sciences (BES) through the Stanford Synchrotron Radiation Lightsource (SSRL), and the Swedish Research Council for financial support. The authors also acknowledge the technical support from B. Johnson, V. Borzenets, D. Van Campen and R. Marks in the SSRL x-ray support group. Finally, the authors thank F. Sciortino, L. Fu, A. Mehta and S. Brennan for helpful discussions. The MD simulations were performed on resources provided by the Swedish National Infrastructure for Computing (SNIC) at High-Performance Computing Center North (HPC2N).

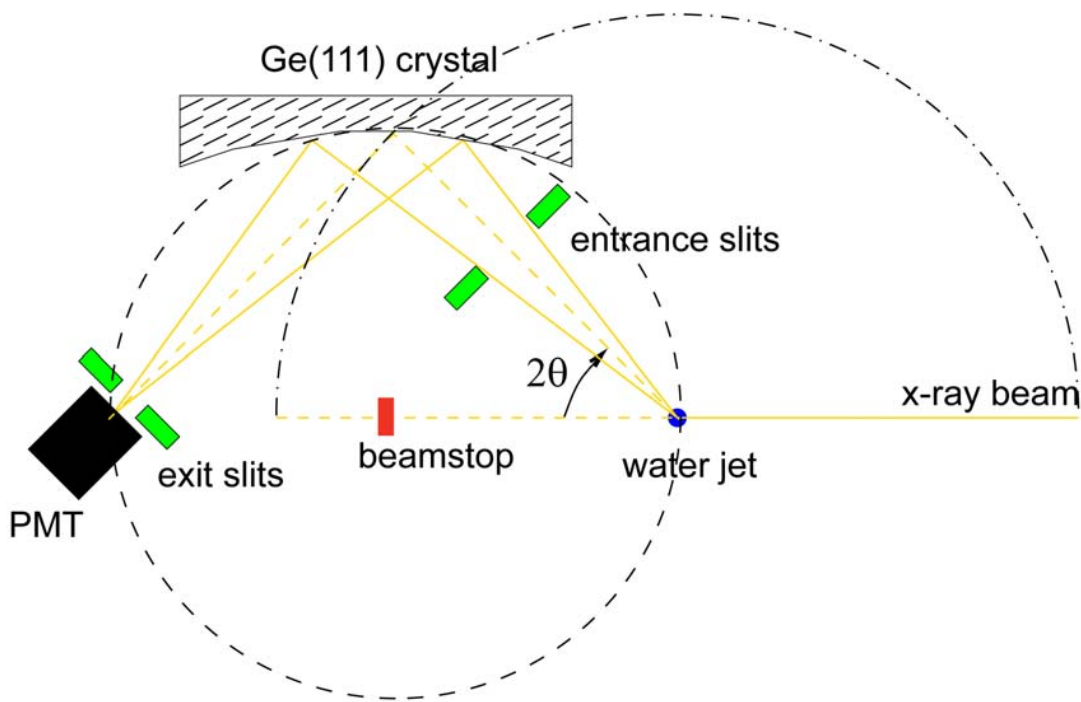

**Figure 1.** Scattering geometry of the current XRD study of water structure. A container-free water jet with 360 μm in diameter is aligned in the rotation center of 6-circle diffractometer. A spherically curved Ge(111) analyzer crystal, satisfying the Rowland scattering geometry at 17 keV, is mounted with its center always on the rotation trace of 2θ scattering angle. A PMT detector is mounted at the focus point of the Rowland circle and its chance measuring the incident beam is eliminated by a beamstop right after the water jet, leading to a minimum accessible $q$ value of 0.5 Å$^{-1}$. The maximum $q$ value of 16 Å$^{-1}$ is determined by the maximum 2θ of ~135 °.

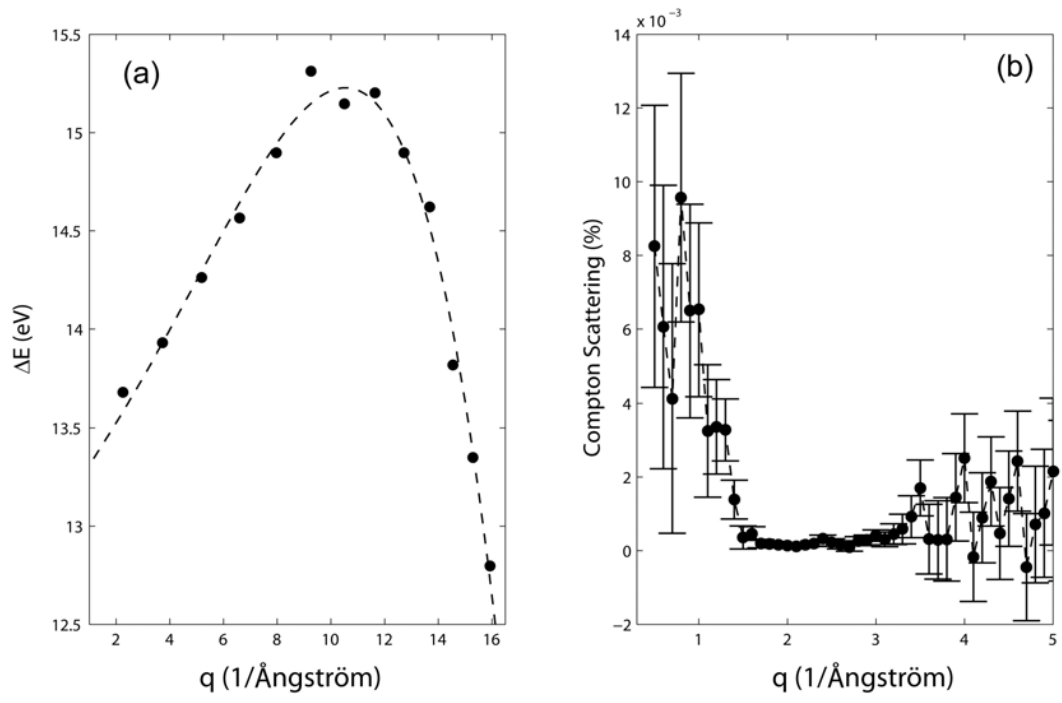

**Figure 2.** The *q*-dependent (**a**) energy resolution and (**b**) Compton scattering contribution, measured at ~30 eV away from the elastic peak at low *q*.

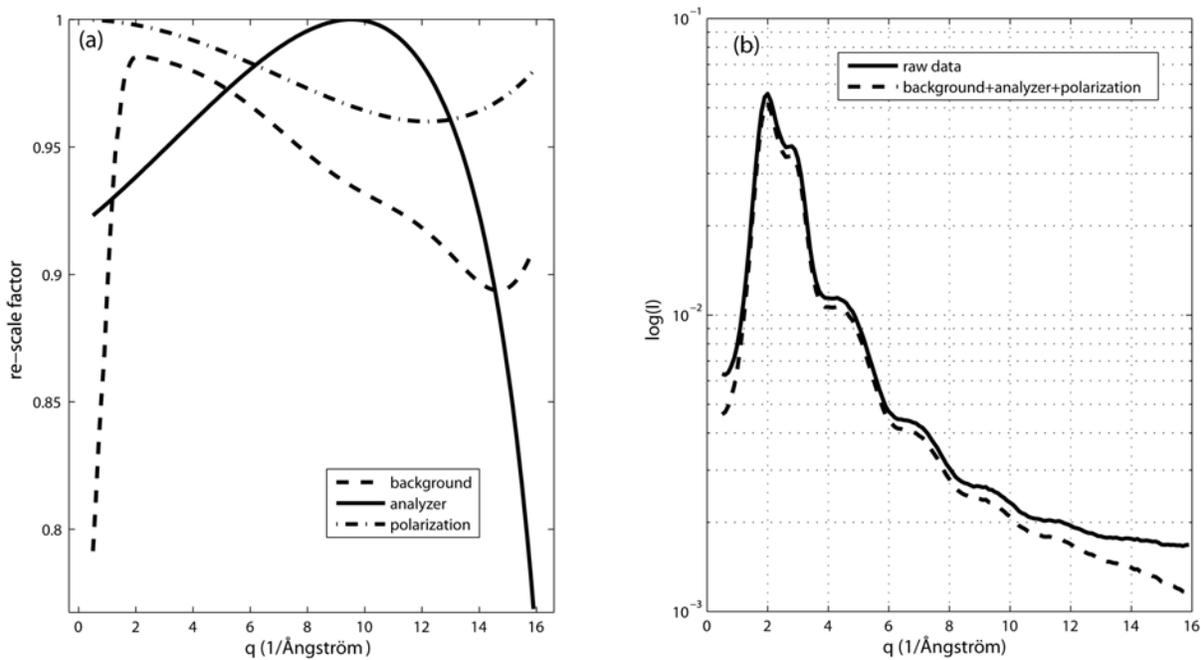

**Figure 3.** (**a**) Re-scale factors of measured scattering intensity due to the contributions from air scattering background (dashed line), the optical aberration of Ge analyzer crystal (solid line), and the polarization of the incident x-ray beam (dash-dotted line). (**b**) Comparison between the raw data of scattering intensity (solid line) and the data after the three corrections (dashed line) shown in (a) with a logarithmic scale for liquid water at 7 °C.

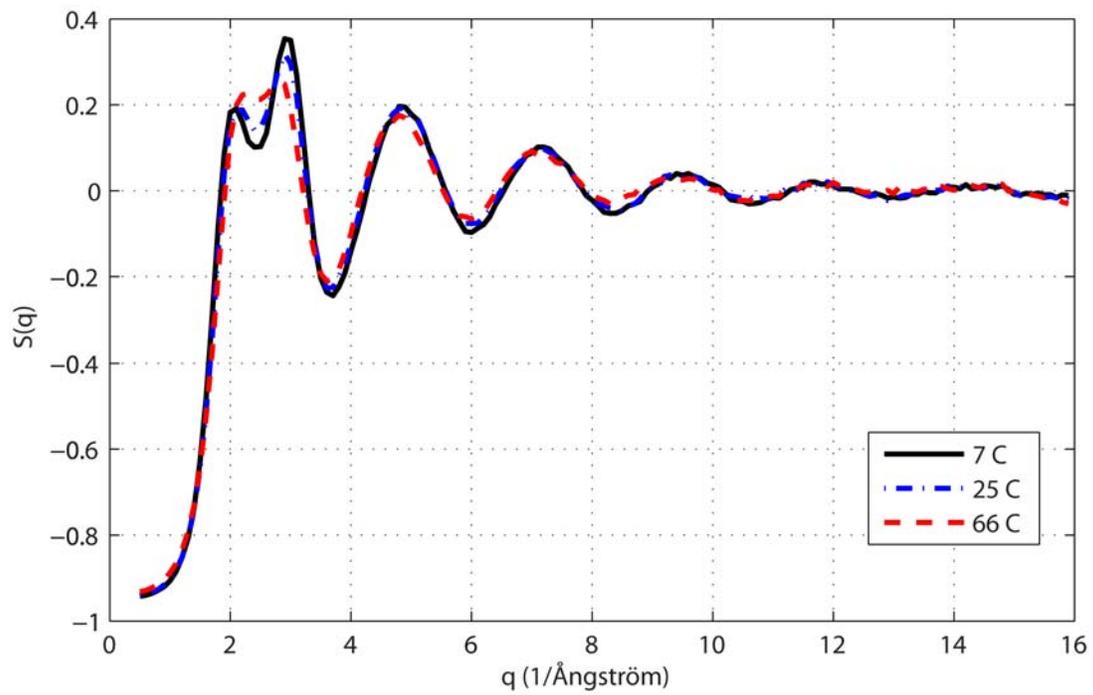

**Figure 4.** Comparison of $S(q)$ of liquid water measured at 7 (black, solid line), 25 (blue, dash-dotted line) and 66 °C (red, dashed line) respectively.

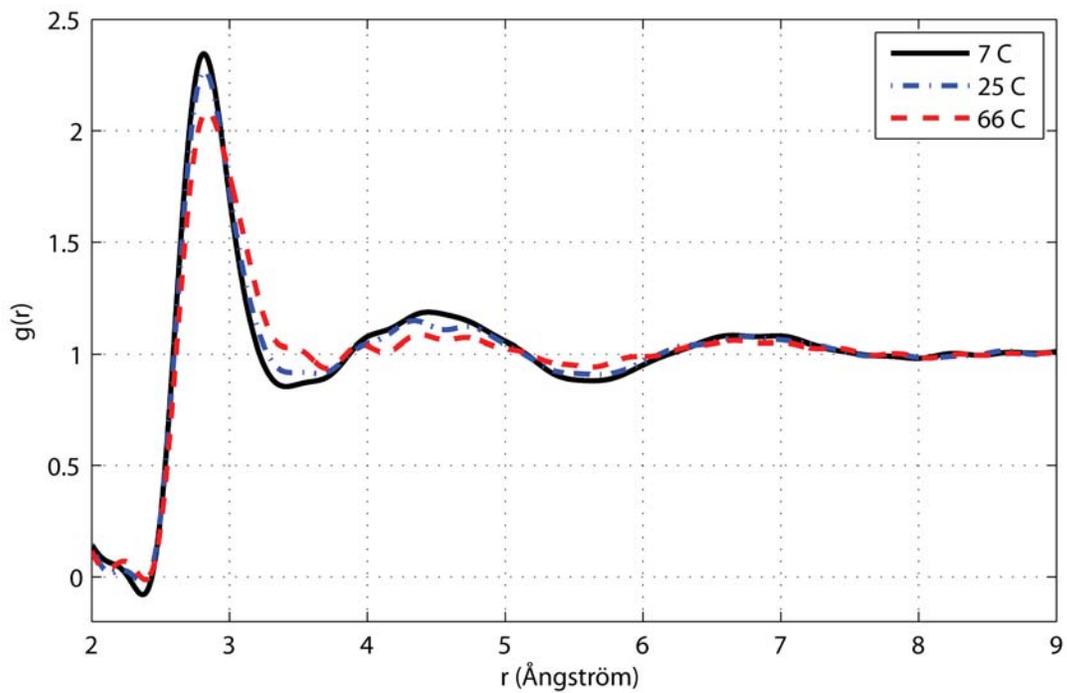

**Figure 5.** Comparison of *g*(*r*) of liquid water measured at 7 (black, solid line), 25 (blue, dash-dotted line) and 66 °C (red, dashed line) respectively.

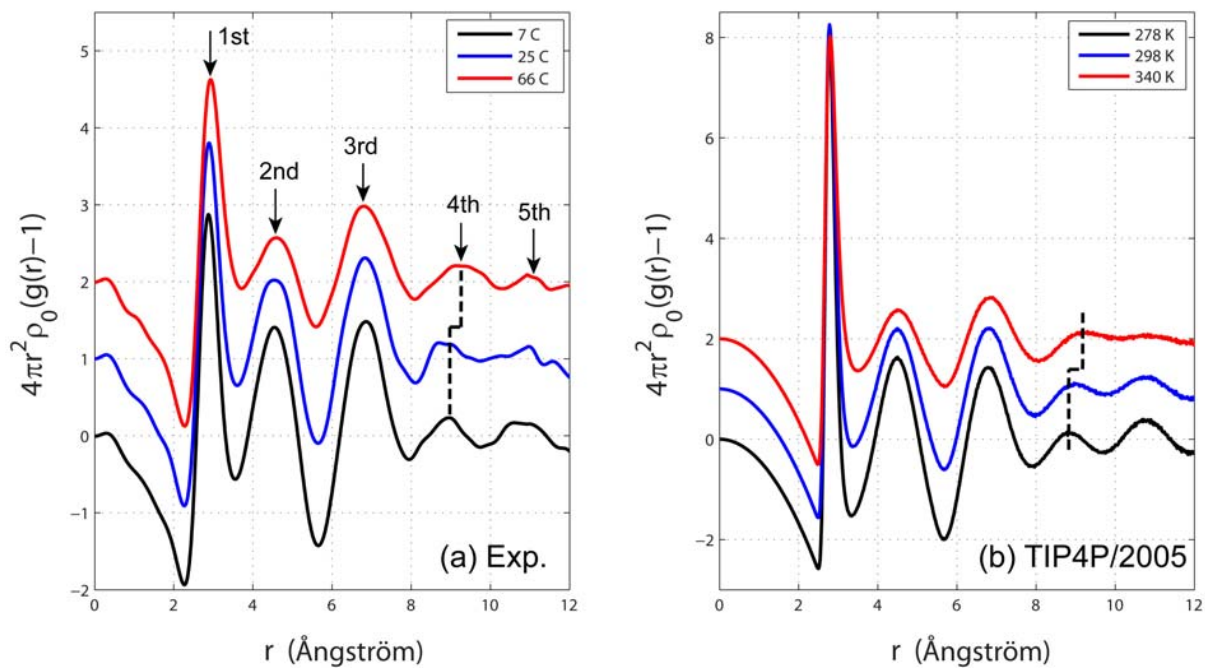

**Figure 6.** Comparison of the dRDFs derived from (**a**) the current XRD measurements with (**b**) a MD simulation using the TIP4P/2005 force field. For clarity, the dRDFs are shifted vertically and temperature decreases from top to bottom as labeled. The locations of the five structure peaks resolved from the experiment are marked by arrows in (**a**). Dashed lines are drawn to indicate the shift of the 4$^{th}$ peak in both experiment and simulation.

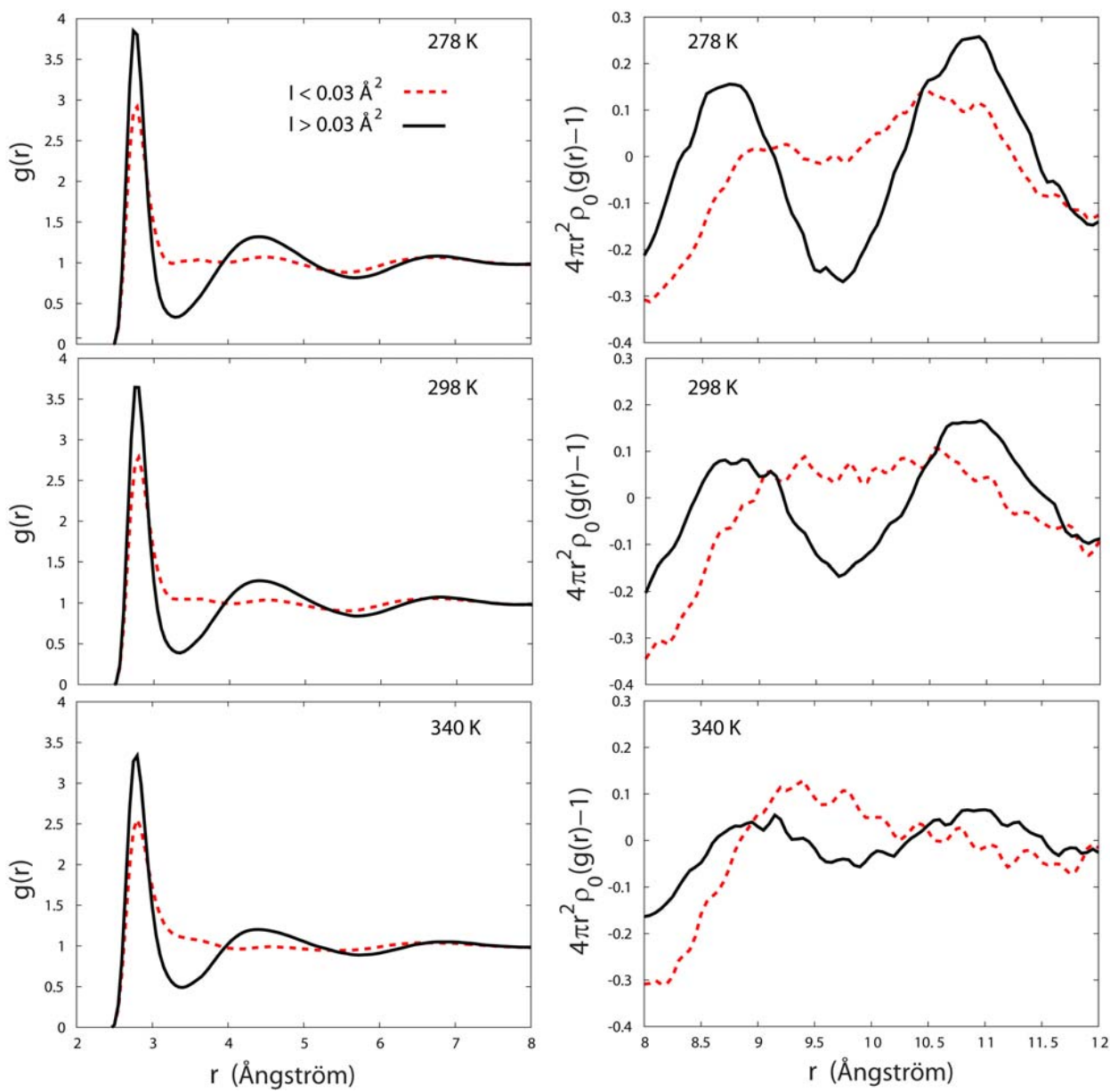

**Figure 7.** Decomposed contributions to the (left) O-O PCFs and (right) O-O dRDFs, based on decomposing TIP4P/2005 trajectories into sub-species defined according to the local-structure index $I(i)$. The applied cut-off value is $I_c$=0.03 Å$^2$, resulting in relative populations of high-LSI species of 49%, 44% and 38% at 278 K, 298 K and 340 K, respectively. The dRDFs have been scaled by these fractions to reflect the relative contributions.